\documentclass{article}
\usepackage{spconf,amsmath,graphicx}

\usepackage{siunitx}
\usepackage{psfrag}
\usepackage{paralist}
\usepackage{tikz}
\usepackage{pgfplots}
\usepackage{microtype}
\usepackage{amssymb}
\usepackage[]{hyperref}

\DeclareMathOperator*{\argmin}{argmin}
\DeclareMathOperator*{\argmax}{argmax}

\newcommand{\Eqref}[1]{Eq.~\eqref{#1}}
\newcommand{\Figref}[1]{Fig.~\ref{#1}}
\newcommand{\Secref}[1]{Sec.~\ref{#1}}
\newcommand{\Tabref}[1]{Tab.~\ref{#1}}

\newlength{\fheight}
\newlength{\fwidth}

\title{Absolute Geometry Calibration of Distributed Microphone Arrays\\ in an Audio-Visual Sensor Network}
\name{Florian Jacob, Reinhold Haeb-Umbach\thanks{This work has been supported by Deutsche Forschungsgemeinschaft (DFG) under contract no. Ha3455/7-2.}}
\address{Department of Communications Engineering, University of Paderborn, Germany\\
\small \tt \{jacob, haeb\}@nt.uni-paderborn.de}

\begin{document}
\maketitle
\begin{abstract}

Joint audio-visual speaker tracking requires that the locations of microphones and cameras are known and that they are given in a common coordinate system. Sensor self-localization algorithms, however, are usually separately developed for either the acoustic or the visual modality and return their positions in a modality specific coordinate system, often with an unknown rotation, scaling and translation  between the two.
In this paper we propose two techniques to determine the positions of acoustic sensors in a common coordinate system, based on audio-visual correlates, i.e., events that are localized by both, microphones and cameras separately. The first approach maps the output of an acoustic self-calibration algorithm by estimating rotation, scale and translation to the visual coordinate system, while the second solves a joint system of equations with acoustic and visual directions of arrival as input. The evaluation of the two strategies reveals that joint calibration outperforms the mapping approach and achieves an overall calibration error of $\SI{0.20}{m}$ even in reverberant environments.
\end{abstract}
\section{Introduction}
\label{sec:intro}
Advanced teleconferencing systems, smart rooms or surveillance and monitoring systems are example applications of distributed audio-visual sensor networks. For many tasks, such as automatic camera steering, events or objects of interest have to be localized either acoustically, visually or jointly, which in turn requires that the positions of the sensors need to be known. While the sensor positions can be determined manually, it is more convenient to do so automatically, in particular if they can change over time, e.g., if a smartphone, which is part of the network, is carried by a moving person.

Automatic geometry calibration of sensors is typically realized by localizing and tracking an object and subsequently determining the position of the sensors, such that the measurements of the object's positions are most plausible.

Visual calibration algorithms work on features extracted from the camera images. They can be divided into two categories \cite{Bru2011}. The first one tries to extract these features from easily recognizable objects \cite{Chen00widearea}, whereas the second group extracts features from an arbitrary scene to compare the field of view of the individual cameras \cite{Bruckner10:ASM}.

For acoustic sensor nodes time of flight (ToF) based algorithms, which employ special calibration hardware and signals to achieve high positioning accuracies \cite{1369311,Crocco:etal:2012}, have been proposed. However, a tight clock synchronization between transmitter and receiver is required, whereas \cite{6637618} relaxed this limitation by estimating the differences in the sampling phase and the sensor positions jointly. If the calibration is based on time difference of arrival (TDoA) measurements, loudspeaker and microphones need no longer be synchronized, and a human speaker can be used as sound source. However, a clock synchronisation of the A/D converters of the distributed microphones is still required. Even this requirement becomes obsolete if direction of arrival (DoA) based techniques are employed \cite{JaScHa12,Jacob2013}. TDoA and DoA based calibration if carried out with artificial calibration signals with appropriate correlation properties, will typically achieve higher accuracy compared to speech signal based approaches \cite{5661986}. Calibration based on natural speech is preferable from a usability point of view, as it can be carried out in the background unnoticed by the users of the audio-visual sensor network.

Most geometry calibration techniques 
are unable to report the sensor positions in absolute coordinates. They 
return their estimates in a modality specific coordinate system, resulting in an unknown  rotation,  translation and scaling between the coordinate axes of the acoustic and visual sensor network. The scaling ambiguity can be fixed if ToA or TDoA measurements are employed \cite{Hennecke2011-TAS,ScJaHaHeFi11}. If the calibration is based solely on DoA measurements, the scale ambiguity still remains, regardless of the modality used \cite{ScJaHaHeFi11, 5206810}.

If the sensor positions of one modality are known, the displacement between the coordinate systems can be resolved by exploiting audio-visual correlates, i.e., events or objects that can be localized both acoustically and visually \cite{JaHa2014,PlingAV}. In this paper we build upon this idea and present two strategies to localize the acoustic sensors in a joint audio-visual coordinate system. Both the acoustic and visual localization is solely based on DoA measurements as they impose the least synchronisation requirements as detailed above.

The first approach uses the existing acoustic sensor calibration techniques from \cite{JaScHa12}. Based on the relative geometry estimates the speaker trajectory can be recovered with the intersection based approach from \cite{479481}, while simultaneously the speaker trajectory is estimated by the visual sensor network. By computing the optimal mapping between the acoustic and visual trajectory we are able to reveal rotation, translation and scale between both modalities. The second approach exploits the fact that the sensors of both modalities deliver DoA estimates. Thus, acoustic and visual measurements can be cast in a single system of equations to determine the acoustic sensor positions, while the known visual sensor positions serve as anchor positions to eliminate the scale ambiguity. A key component of both DoA based calibration methods is the random sample consensus (RANSAC) outlier rejection algorithm \cite{RanSaC81}, which diminishes the impact of poor DoA estimates on the localization performance. In case of the joint calibration, this scheme will not only reject acoustic DoA outliers, it will reject visual DoA outliers as well.

In the next section we introduce the first approach based on a coordinate mapping, whereas \Secref{sec:calibration} describes the joint calibration approach. The performance of both algorithms is evaluated in \Secref{sec:results}, before \Secref{sec:conclusion} concludes this paper.

\section{Coordinate Mapping}
\label{sec:mapping}
Our goal is the estimation of the coordinates of $I$ acoustic sensors, where the coordinate system is defined by the known positions of $K$ visual sensors. The location of the $k$-th visual sensor node is described in 2D by the position vector $\mathbf{c}_k$ and  orientation $\gamma_k$. Now, consider a moving speaker located at position $\mathbf{e}_t$ at time $t$, who is seen by the visual sensors at DoAs $\delta_{k,t}$, $k{=}1, \dots, K$. A position estimate $\mathbf{e}_t$ is obtained from the DoAs by the intersection based technique presented in \cite{479481}.

The acoustic DoA estimates $\varphi_{i,t}$, $i{=}1,\dots , I$; $t{=}1, \dots, T$, captured from the same speaker trajectory are used to determine estimates $\tilde{\mathbf{m}}_i$,  $i{=}1,\cdots,I$, of the acoustic sensor positions and estimates $\Theta_i$ of the orientations, using the calibration algorithm from \cite{JaScHa12}.
This algorithm can only provide a relative geometry with an unknown scale factor. Therefore, only relative speaker position estimates $\tilde{\mathbf{e}}_t$ are obtained, using the same intersection based method as above.
 
Since the acoustic event locations $\tilde{\mathbf{e}}_t$ are described in a different coordinate system as the visual estimates $\mathbf{e}_t$, there arises the following coordinate mapping problem:
\begin{align}
  \label{eq:coordinateTransformation}
  \mathbf{e}_t = s\mathbf{R}\tilde{\mathbf{e}}_t+\mathbf{d} \text{;\quad}  t=1, \dots , T \text{,}
\end{align}
where $s$ models the unknown scale factor and $\mathbf{R}$ and $\mathbf{d}$ the rotation and translation between the acoustic and the visual coordinate system. Mapping a set of points from one coordinate system to another is known as Rigid Body Transformation (RBT). In contrast to the widespread approach from \cite{Challis} to compute the RBT parameters (scale, rotation and translation) via a Singular Value Decomposition (SVD) we suggest a computation in the Discrete Fourier transform (DFT) domain, which turned out to be computationally more efficient.

Hence, we introduce a complex representation of the estimated speaker positions as $u_t{=}\tilde{e}_{1,t}{+}\mathrm{j}\tilde{e}_{2,t}$ and  $v_t{=}e_{1,t}{+}\mathrm{j}e_{2,t}$ respectively, where $\tilde{\mathbf{e}}_t {=} (\tilde{e}_{1,t}, \tilde{e}_{2,t})$ and $\mathbf{e}_t {=} (e_{1,t}, e_{2,t})$ are the two-dimensional speaker positions in the acoustic and visual coordinate system, respectively. Thus, the mapping problem of \Eqref{eq:coordinateTransformation} is expressed as
\begin{align}
  \label{eq:shapeMapping}
  v_t = \alpha u_t + \beta\text{;\ } \alpha,\beta\in\mathbb{C}\text{.}
\end{align}
The absolute value and the phase of $\alpha$ correspond to scale and orientation, while $\beta$ corresponds to the translation. Arranging all observations into vectors $\mathbf{v} {=} [v_1,\dots,v_T]^\mathrm{T}$ and $\mathbf{u} {=} [u_1,\dots,u_T]^\mathrm{T}$ the least squares estimate for the RBT parameters in the complex space is given by
\begin{align}
 \label{eq:objComplex}
 \langle \alpha^\ast, \beta^\ast \rangle = \underset{\alpha,\beta}{\argmin} \left(\alpha \mathbf{u} + \beta\mathbf{1} - \mathbf{v}\right)^\mathrm{H}\left(\alpha \mathbf{u} + \beta\mathbf{1} - \mathbf{v}\right) \text{,}
\end{align}
where $\mathbf{1}$ denotes an $T$-element vector of ones and $(\cdot)^\mathrm{H}$ the complex conjugate transpose of a vector.

Let  $\mathbf{x}$ and $\mathbf{y}$ denote the DFTs of $\mathbf{u}$ and $\mathbf{v}$. The optimization problem of \Eqref{eq:objComplex} is expressed in the DFT domain as
\begin{align}
 \langle \alpha^\ast, \beta^\ast \rangle = \underset{\alpha,\beta}{\argmin} \left(\alpha \mathbf{x} + \beta\mathbf{z} - \mathbf{y}\right)^\mathrm{H}\left(\alpha \mathbf{x} + \beta\mathbf{z} - \mathbf{y}\right) \text{,}
\end{align}
where $\mathbf{z} {=}\left[\begin{matrix}1,0,\dots,0\end{matrix}\right]^\text{T}$ is a vector of length $T$.
Due to the orthogonality properties of the DFT the joint optimization is decoupled into two separate optimizations:
\begin{align}
  \label{eq:objShape}
   \alpha^\ast &= \underset{\alpha}{\argmin} \left(\alpha\mathbf{x}_{2:T}{-}\mathbf{y}_{2:T}\right)^\mathrm{H}\left(\alpha\mathbf{x}_{2:T}{-}\mathbf{y}_{2:T}\right)\text{ and}
\end{align}
\vspace*{-0.8cm}
\begin{align}
  \label{eq:obj3}
  \beta^\ast &= \underset{\beta}{\argmin} \left(\alpha^\ast x_{1}{+}\beta{-}y_{1}\right)^\mathrm{H}\left(\alpha^\ast x_{1}{+}\beta-y_{1}\right)\text{,}
\end{align}
where the first bin of the DFTs is denoted by $(\cdot)_1$ and all other bins by $(\cdot)_{2:T}$. Since \Eqref{eq:objShape} and \Eqref{eq:obj3} are general least squares problems, the solution is found to be
\begin{align}
 \label{eq:AlphaBeta}
 \alpha^\ast = \mathbf{x}_{2:T}^\mathrm{H}\mathbf{y}_{2:T} / \left(\mathbf{x}_{2:T}^\mathrm{H}\mathbf{x}_{2:T}\right)\text{\ and \ }
 \beta^\ast = {y}_1-\alpha^\ast {x}_1\text{.}
\end{align}

The RBT parameters  can be retrieved  as follows:
\begin{align}
  \label{eq:RBT}
  s = \left|\alpha\right|\text{, }
  \mathbf{R} =
  \left[\begin{smallmatrix}
    \Re\left\{\frac{\alpha}{s}\right\}& -\Im\left\{\frac{\alpha}{s}\right\}\\
    \Im\left\{\frac{\alpha}{s}\right\} & \Re\left\{\frac{\alpha}{s}\right\}
  \end{smallmatrix}\right]\text{ and }
  \mathbf{d} =
  \frac{1}{N}\left[\begin{smallmatrix}
    \Re\left\{\beta\right\} \\\Im\left\{\beta\right\}
  \end{smallmatrix}\right]\text{,}
\end{align}
where $\Re$ and $\Im$ denote real and imaginary part, respectively. If this transformation is applied to the relative acoustic sensor position estimate $\tilde{\mathbf{m}}_i$ according to \Eqref{eq:coordinateTransformation}, the absolute acoustic sensor positions $\mathbf{m}_i$ in the visual coordinate system are obtained.

To summarize, the calibration algorithm to recover the acoustic sensor positions in the visual coordinate system consists of three steps. First, run the relative acoustic calibration algorithm and estimate the speaker trajectory. At the same time, track the speaker in the visual domain. Secondly, compute the DFTs of both trajectories, evaluate \Eqref{eq:AlphaBeta} and compute the RBT parameters by \Eqref{eq:RBT}. Finally, use the RBT parameters to transform the acoustic sensor position estimates  from the first step into the visual coordinate system.

The DFT based RBT parameter estimation delivers the same results as the conventional SVD based technique \cite{Challis}, but our FFT based implementation is twice as fast as the SVD.

\section{Joint Calibration}
\label{sec:calibration}

Since the acoustic and the visual sensors deliver DoA estimates, we propose to extend the calibration algorithm that was used for the acoustic sensors only in step one of the algorithm presented in the last section, to both modalities and jointly calibrate the audio-visual network. Due to the known positions of the visual sensors, the scale ambiguity vanishes.

\begin{figure}[b]
 \centering
 \psfrag{e}{$\mathbf{e}_{t}$}
 \psfrag{p}{$\mathbf{m}_i$}
 \psfrag{ph}{$\varphi_{i,t}$}
 \psfrag{k}[cB][lb]{$\Theta_i$}
 \psfrag{a}{$\widehat{\varphi}_{i,t}$}
 \psfrag{xp}[c][c]{$m_{1,i}$}
 \psfrag{xe}[c][c]{$e_{1,t}$}
 \psfrag{yp}[cr][cc]{$m_{2,i}$}
 \psfrag{ye}[cr][cc]{$e_{2,t}$}
 \psfrag{x}{$x$}
 \psfrag{y}{$y$}
 \includegraphics[width=0.5\linewidth]{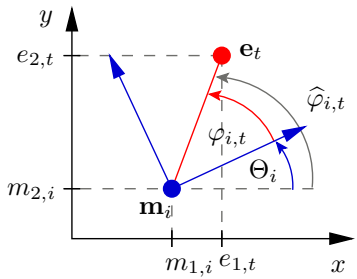}
 \vspace{-0.27cm}
 \caption{Geometric relation between acoustic sensor and event location.}
 \label{fig:DoA}
\end{figure}

In the local coordinate system of the $i$-th acoustic sensor a DoA measurement can be modelled as a unit length vector
\begin{align}
 \mathbf{f}_{i,t} = [\begin{matrix}\cos\left(\varphi_{i,t}\right) & \sin\left(\varphi_{i,t}\right)\end{matrix} ]^\mathrm{T} \text{,}
\end{align}
pointing from the sensor position to the event location. This measurement vector will be compared with a prediction vector
\begin{align} \label{Eq:vp}
  \widehat{\mathbf{f}}_{i,t}
  &=
  \left[
    \begin{matrix}
     \cos \left( \widehat{\varphi}_{i,t}-\Theta_i \right) \\
     \sin \left( \widehat{\varphi}_{i,t}-\Theta_i \right)
    \end{matrix}
  \right] \text{,}
\end{align}
where $\widehat{\varphi}_{i,t} {=} \arg\left\{\mathbf{e}_{t}-\mathbf{m}_i\right\}$, see \Figref{fig:DoA}. Following our previous publication \cite{JaScHa12} this prediction can be formulated as a function of the geometry parameters as follows:
\begin{align}
 \widehat{\mathbf{f}}_{i,t}
  &=
  \left[
    \begin{matrix}
     \cos(\Theta_i) & \sin(\Theta_i) \\
     -\sin(\Theta_i) & \cos(\Theta_i)
    \end{matrix}
  \right]
  \frac{\mathbf{e}_{t}-\mathbf{m}_i}{\|\mathbf{e}_{t}-\mathbf{m}_i\|} \text{.}
\end{align}

By introducing the abbreviation
\begin{align}
  f = \sum\limits_{i=1}^I\sum\limits_{t=1}^T \|\mathbf{f}_{i,t}^\mathrm{T}\widehat{\mathbf{f}}_{i,t}\|^2
\end{align}
and arranging the sensor positions, sensor orientations and events into matrices $\mathbf{M} {=} [\mathbf{m}_1, \dots, \mathbf{m}_I]$, $\mathbf{\Theta} {=} [\Theta_1,\dots, \Theta_I]$ and $\mathbf{E} {=} [\mathbf{e}_1,\dots, \mathbf{e}_T]$ respectively, the geometry can be recovered by
\begin{align}
 \label{eq:OptCalib}
 \langle \mathbf{M}^*, \mathbf{\Theta^*},\mathbf{E}^* \rangle = \underset{\mathbf{M},\mathbf{\Theta},\mathbf{E}}{\argmax}\left\{f\right\} \text{.}
\end{align}
The maximization problem of \Eqref{eq:OptCalib} can easily be transformed into a root-finding problem, since $\mathbf{f}_{i,t}$ and $\widehat{\mathbf{f}}_{i,t}$ are unit length vectors. Subsequently, the minimization is carried out by Newton's method.

The formulation for the estimated and predicted DoA vectors hold for the visual sensors, too. Thus we define
\begin{align}
 \mathbf{g}_{k,t} &= [\begin{matrix}\cos\left(\delta_{k,t}\right) & \sin\left(\delta_{k,t}\right)\end{matrix} ]^\mathrm{T} \text{ and } \\
 \widehat{\mathbf{g}}_{k,t}
  &=
  \left[
    \begin{matrix}
     \cos(\gamma_k) & \sin(\gamma_k) \\
     -\sin(\gamma_k) & \cos(\gamma_k)
    \end{matrix}
  \right]
  \frac{\mathbf{e}_{t}-\mathbf{c}_k}{\|\mathbf{e}_{t}-\mathbf{c}_k\|} \text{,}
\end{align}
with the only difference that the visual sensor positions $\mathbf{c}_k$ and the corresponding orientations $\gamma_k$ are known. Hence, the visual DoA measurements form additional constraints for the optimization of \Eqref{eq:OptCalib}, and we incorporate them to obtain a formulation which allows a joint audio-visual calibration:
\begin{align}
 \label{eq:OptCalibNew}
 \langle \mathbf{M}^*, \mathbf{\Theta^*},\mathbf{E}^* \rangle {=} \underset{\mathbf{M},\mathbf{\Theta},\mathbf{E}}{\argmax}\left\{f + \sum\limits_{k=1}^K \sum\limits_{t=1}^T \|\mathbf{g}_{k,t}^\mathrm{T}\widehat{\mathbf{g}}_{k,t}\|^2 \right\} \text{.}
\end{align}
The optimization of \Eqref{eq:OptCalibNew} is again turned into a root-finding problem in order to apply Newton's method, where the visual measurements provide the required constraints to obtain an absolute sensor position estimate in the coordinate system defined by the visual sensors.

In the noise free case, with perfect DoA measurements, the sensor positions and orientations can perfectly be recovered, but imperfect acoustic or visual DoA estimates caused by reverberation or false detections can prevent a successful optimization. Our earlier investigations presented in \cite{JaScHa12} showed, that this issue can successfully be addressed by the RANSAC \cite{RanSaC81}. Since the application of the RANSAC is straightforward we highlight only the relevant parts. The procedure can be summarized as follows:
\begin{compactenum}
 \item Randomly select the minimal number of observations necessary to solve \Eqref{eq:OptCalibNew}, e.g. $T>\frac{3I}{I+K-2}$.
 \item Determine sensor positions and orientations based on the selected observations by solving \Eqref{eq:OptCalibNew}.
 \item Compute the intersection of all DoA axes for each event. The hypothesized event location is the mean of all intersections. A DoA measured by a sensor becomes part of the candidate set $\mathcal{C}$, if the average distance of all its intersection points to the hypothesized event location is smaller than a threshold.
 \item If the number of elements in $\mathcal{C}$ is larger than the consensus set, estimate the sensor positions and orientations based on $\mathcal{C}$. It becomes the new consensus if its error is smaller than the error of the current consensus.
 \item If the number of elements in $\mathcal{C}$ is smaller than consensus set, choose a new initial set or stop the algorithm as soon as the maximum number of iterations is reached.
\end{compactenum}
As a modification of this standard approach, we used the updated consensus set of step 4 as the input for the second step.

\section{Simulation results}
\label{sec:results}
In order to evaluate the performance of both calibration strategies we used the following simulation framework. We simulated $3$ random speaker trajectories, where the speaker stops at approximately $140$ positions for $\SI{5}{seconds}$ before he moves on. The sensors are located in a room of size $\SI{6.2}{m} \times \SI{7.2}{m}$. $4$ simulated cameras and $4$ simulated five-element circular microphone arrays (radius $\SI{5}{cm}$) are located sufficiently far apart from the walls, where the cameras were oriented towards the center of the room. The microphone signals are generated by the Image Method \cite{Imag}, for reverberation times from $\SI{0}{ms}$ up to $\SI{500}{ms}$. Acoustic DoA estimates are obtained by correlating the filter impulse responses of a filter-and-sum beamformer, which continuously adapts to the moving source \cite{War05}.

Rather than working on a true camera signal, visual DoA estimates are simulated as follows. We employ Hidden Markov Models (HMMs) to describe the errors in the DoA estimation. A limited field of view of the camera is taken into account by dropping all angles outside a window of $\pm\SI{30}{\degree}$ relative to the camera orientation. This effect is modelled by two separate HMMs. The first HMM is for the case that a speaker is inside the visible region of the camera. Here we distinguish the states 'detection', 'missed detection' and 'false detection'. The second HMM models the case that no speaker is inside the visible region. It incorporates the states 'false detection' and 'no detection'.
The transition probabilities of these models and the variance of the error distribution have been learned by computing  histograms of oriented gradients (HOG) and applying a support vector machine (SVM) to identify the head and shoulder region of the speaker on the AV16.3 audio-visual corpus \cite{lathoud04c}, using the annotated sequences \textit{seq01-1p-0000} and \textit{seq15-1p-0100}.

In order to perform a fair comparison between the approaches presented in \Secref{sec:mapping} and \Secref{sec:calibration} the estimation of the RBT parameters is embedded into a RANSAC framework, too, since we have shown in \cite{JaHa2014} that the RANSAC can boost the performance of the estimation of the RBT parameters.

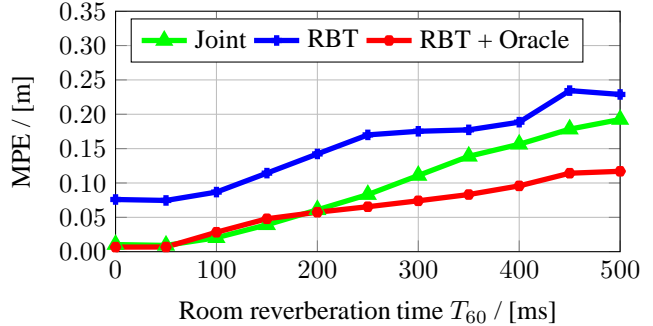
\begin{figure}
\begin{center}
\setlength{\fheight}{3.2cm}
\setlength{\fwidth}{0.78\linewidth}
%
%
\begin{tikzpicture}

\begin{axis}[%
width=\fwidth,
height=\fheight,
scale only axis,
xmin=0,
xmax=500,
xtick={  0, 100, 200, 300, 400, 500},
xlabel={Room reverberation time $T_{60}$ / [ms]},
xmajorgrids,
ymin=0,
ymax=0.35,
ytick={   0, 0.05,  0.1, 0.15,  0.2, 0.25,  0.3, 0.35},
ylabel={ MPE / [m]},
ymajorgrids,
y tick label style={
        /pgf/number format/.cd,
            fixed,
            fixed zerofill,
            precision=2,
        /tikz/.cd
    },
legend style={at={(0.03,0.97)},anchor=north west,legend columns=3,draw=black,fill=white,legend cell align=left}
]
\addplot [color=green,solid,line width=2.0pt,mark=triangle,mark options={solid}]
  table[row sep=crcr]{0	0.0103655980701472\\
50	0.00918626525696592\\
100	0.0203554751476251\\
150	0.0391610666906216\\
200	0.0609712852879601\\
250	0.0830872386009031\\
300	0.110898328330095\\
350	0.138816310990919\\
400	0.156338569582906\\
450	0.178336302649358\\
500	0.192390588162816\\
};
\addlegendentry{Joint};

\addplot [color=blue,solid,line width=2.0pt,mark=+,mark options={solid}]
  table[row sep=crcr]{0	0.0759752251842278\\
50	0.074591151854937\\
100	0.0867627276128565\\
150	0.114359155331134\\
200	0.142268313000633\\
250	0.170005022559309\\
300	0.175243091571081\\
350	0.177316269450347\\
400	0.188504987550138\\
450	0.234432400857396\\
500	0.22883676786828\\
};
\addlegendentry{RBT};

\addplot [color=red,solid,line width=2.0pt,mark=asterisk,mark options={solid}]
  table[row sep=crcr]{0	0.00671517756014653\\
50	0.00685031837885226\\
100	0.0283147096183855\\
150	0.0479366899942488\\
200	0.0574643864891903\\
250	0.0654519345478461\\
300	0.0740736787848106\\
350	0.0831938441883096\\
400	0.0958059348479508\\
450	0.11422893668533\\
500	0.117008658439493\\
};
\addlegendentry{RBT + Oracle};

\end{axis}
\end{tikzpicture}%
\end{center}
\vspace*{-0.8cm}
\caption{Comparison of mean positioning error (MPE) for joint audio-visual calibration (joint), calibration by coordinate mapping (RBT) and  coordinate mapping with an oracle information (RBT + oracle).}
\label{fig:err}
\vspace*{-0.15cm}
\end{figure}

Since the RANSAC is a random process, we average over multiple runs. A sensor configuration is characterized by the positions and orientations. \Figref{fig:err} compares the mean positioning error (MPE) of the coordinate mapping based calibration (RBT) and the joint calibration strategy (Joint). It can be observed that the joint calibration clearly outperforms the RBT approach, in particular at low reverberation times. Obviously, it is advantageous to avoid premature decisions on acoustic source and sensor positions until the visual information is accounted for, as it is done in the joint calibration approach.

The coordinate mapping approach has limited capabilities to determine a precise scale factor, and errors in the scale factor dominate its performance. In order to isolate scale factor estimation errors from orientation and translation errors we performed an oracle experiment, where the scaling  is assumed to be known. Indeed, the performance is now similar to that of the joint approach for low reverberation times and superior in a highly reverberant environment. The sensor orientation error of both approaches is approximately the same and smaller than $\SI{2}{\degree}$ for all reverberation times.

To achieve precise calibration results a suitable spatial event configuration is more important than the total number of available events. Thus, we selected $15$ events with an appropriate configuration of one exemplar trajectory and perform a joint calibration. The results of \Tabref{tab:c1} show that a similar performance as in the previous experiment, which used the complete trajectory, is possible.

\begin{table}[h]
\centering
\begin{tabular}{l|c|c|c|c|c|c}
$T_{60}$ / ms & 0 & 100 & 200 & 300 & 400 & 500\\\hline
MPE / $\mathrm{m}$     & 0.01 & 0.02 & 0.06 & 0.14 & 0.13 & 0.23
\end{tabular}
\caption{Joint calibration using 15 events with appropriate spatial configuration.}
\label{tab:c1}
\vspace*{-0.05cm}
\end{table}

\section{Conclusions}
\label{sec:conclusion}
We have described two different strategies to obtain an absolute calibration of an acoustic sensor network if it is combined with a visual sensor network, whose sensor positions are known. By using one of the two strategies, the scaling problem identified in earlier publications \cite{Bruckner10:ASM, JaScHa12, ScJaHaHeFi11} can be solved. The first approach, which relies on the mapping of an acoustic to a visual speaker trajectory, works with arbitrary acoustic calibration strategies and is therefore very flexible. However, the performance is limited due to the scale estimation errors. The second approach, which is based on the solution of a system of nonlinear equations employing acoustic and visual DoA measurements, is computationally more complex. It outperformed the first approach for all reverberation times and delivered a calibration error smaller than $\SI{0.20}{m}$ and $\SI{2}{\degree}$ even in reverberant environments.

\vfill\pagebreak

\bibliographystyle{IEEEbib}
\bibliography{refs}

\end{document}